\documentclass[conference, 9pt]{IEEEtran}

\IEEEoverridecommandlockouts
\usepackage{cite}
\usepackage{amsmath,amssymb,amsfonts}
\usepackage{algorithmic}
\usepackage{graphicx}
\usepackage{textcomp}
\usepackage{xcolor}
\def\BibTeX{{\rm B\kern-.05em{\sc i\kern-.025em b}\kern-.08em
    T\kern-.1667em\lower.7ex\hbox{E}\kern-.125emX}}

\usepackage{multirow}

\usepackage{makecell}

\begin{document}


\title{Improving End-to-End Speech Recognition for Dysarthric Speech through In-Domain Data Augmentation}
\author{
    \IEEEauthorblockN{Paban Sapkota\IEEEauthorrefmark{1}, Hemant Kumar Kathania\IEEEauthorrefmark{1}, Sudarsana Reddy Kadiri\IEEEauthorrefmark{2}, and Shrikanth Narayanan\IEEEauthorrefmark{2}}
    \\
    \IEEEauthorblockA{\IEEEauthorrefmark{1} Department of Electronics and Communication Engineering, National Institute of Technology Sikkim, India  \\
    Emails: phec230006@nitsikkim.ac.in, hemant.ece@nitsikkim.ac.in}
    \IEEEauthorblockA{\IEEEauthorrefmark{2} Signal Analysis and Interpretation Laboratory (SAIL), University of Southern California, Los Angeles, USA\\
    Emails: skadiri@usc.edu, shri@usc.edu}
}

\maketitle

\begin{abstract}

Dysarthric speech recognition is crucial for facilitating effective communication among individuals with dysarthria. However, accurately recognizing dysarthric speech poses significant challenges due to varying severity levels and limited data availability. In this paper, we explore data augmentation techniques for dysarthric automatic speech recognition (ASR) systems by fine-tuning the End-to-End pre-trained Wav2Vec2 model, with a specific focus on severity levels. To address the challenges of data scarcity and the need for extensive data in fine-tuning pre-trained ASR systems for dysarthric speech, we investigate four prominent data augmentation methods: Speaking-Rate Modification (SRM), Pitch Modification (PM), Formant Modification (FM), and vocal tract Length Perturbation (VTLP), tailored to different aspects of dysarthria. The study uses individually fine-tuned Wav2Vec2 models for each severity class as baseline systems. Additionally, we conducted severity-specific fine-tuning of the ASR model using augmented data. Results demonstrate distinct efficacy patterns for each augmentation technique across severity levels. The best WERs were achieved with SRM ($s$=0.8) for \textit{low} (9.02\%) and \textit{medium} (38.11\%) severities, and with PM ($\tau$=0.8) for \textit{high} severity (55.15\%), reflecting relative improvements of 30.02\%, 16.64\%, and 15.47\%, respectively. These results confirm the effectiveness of the augmentation methods in improving dysarthric ASR performance.


\end{abstract}

%

\section{Introduction}
\label{sec:intro}

Dysarthria is a motor speech disorder caused by impaired muscle control, affecting the coordination, strength, and precision needed for clear speech. This often leads to slurred, slow, or hard-to-understand speech \cite{dysar}. While ASR for typical speech has advanced significantly, especially with the rise of transformer-based self-supervised learning (SSL) models, these advancements present new opportunities to enhance Dysarthric Automatic Speech Recognition (DyASR) \cite{review}.

Dysarthric speech poses significant challenges for ASR due to variability in acoustic properties and articulation across different severity levels. The difficulty in producing clear speech leads to intelligibility issues both within and across speakers, complicating ASR development. Recent advancements in SSL models, like Wav2Vec2 \cite{baevski}, offer promise in improving ASR performance for dysarthric speech. These transformer-based models, which use large amounts of unlabeled data to capture subtle speech variations, could provide a more robust framework for recognizing dysarthric speech, especially when fine-tuned for specific severity levels.

Researchers have explored various strategies to enhance DyASR system performance. To address the challenge of limited atypical speech data, some studies, such as \cite{artificial_data}, have focused on generating artificial data through non-linear speech tempo modification, yielding notable performance improvements. Domain adaptation of pre-trained ASR models has also emerged as a promising solution to data scarcity, as highlighted in \cite{adaptation}. Enhancing speech quality through denoising techniques, converting disordered speech into a form closer to typical speech, was demonstrated in \cite{enhance}.
In a comparative analysis \cite{bonet}, traditional ASR methods were evaluated against neural network-based ASR using features like filter-banks and mel-filter-banks specifically designed for dysarthric speech. Research in \cite{raw_mag} examined raw magnitude spectra-based multi-stream acoustic modeling, achieving performance gains. Furthermore, integrating raw phase-based representations with raw magnitude spectra using single and multi-stream architectures with cascaded convolutional, recurrent, and fully connected layers showed promising results \cite{raw_phase}.
Additionally, a deep neural network model incorporating MFCC-based i-vectors outperformed other models, including convolutional neural networks (CNNs), gated recurrent units (GRUs), and long short-term memory (LSTM) networks \cite{dnn_ivector}. A study in \cite{hu2024self} examined SSL models as feature extractors combined with traditional filterbank features, showing enhanced ASR performance for dysarthric speech. Finally, \cite{wang2024enhancing} demonstrated improvements in pre-trained SSL models like Wav2Vec2 and WavLM using adversarial data augmentation.

To the best of our knowledge, no studies have yet implemented traditional data augmentation techniques for fine-tuning SSL models (Wav2Vec2) for dysarthric speech recognition. In this study, we systematically investigate four widely used data augmentation methods: speaking-rate modification, pitch modification, formant modification, and vocal tract length perturbation. The main contributions of our study include:
\begin{itemize}
 \setlength{\itemsep}{-0.2pt}
\item A systematic investigation of four popular speech data augmentation methods: speaking-rate modification, pitch modification, formant modification, and vocal tract length perturbation -- for fine-tuning the Wav2Vec2 ASR system on dysarthric speech. 
\item A severity-based ASR analysis, where ASR models are trained on dysarthric speech data from one specific severity level  (e.g., \textit{low} severity) and evaluated on data from other severity levels (e.g., \textit{medium}, and \textit{high} severity), allowing for cross-severity generalization assessment.
\item Determination of the optimal modification factor for each augmentation method, tailored to each severity level individually.
\end{itemize}

\section{Explored data augmentation techniques}
\label{sec:pre-trained SSL}

This work explores four speech data augmentation techniques tailored to dysarthric speech: speaking rate modification, pitch modification, formant modification, and vocal tract length perturbation. Speaking rate modification adjusts speech speed to enhance clarity and model comprehension. Pitch modification accounts for tonal variations typical in dysarthric speech. Formant modification refines vowel articulation by enhancing formant frequencies to improve intelligibility. vocal tract length perturbation simulates variations in vocal tract lengths, mimicking the challenges faced by dysarthric speakers to improve understanding across severity levels. 

\subsection{Speaking-rate modification (SRM)}
\label{subsec:SRM}

Speaking-rate modification was achieved using Time Scale Modification (TSM) based on the Real-Time Iterative Spectrogram Inversion with Look-Ahead (RTISI-LA) algorithm \cite{improving}. The scaling factor \texttt{s}, ranging from $0.5 \leq \texttt{s} \leq 2$, was adjusted in increments of 0.1 during experiments. The algorithm employs Short-Time Fourier Transform Modification (STFTM) for audio signal processing \cite{rathod2023noise}. Each frame of the input signal undergoes windowing and is reconstructed iteratively \cite{zhu2007real}. The modified frame length $(L = 256 \cdot \texttt{s})$ and hop size $(S = \frac{L}{4})$ control the STFT process, altering the signal's time scale. The modified audio is then used in further augmentation steps. A key aspect of this approach is the introduction of phase perturbation \cite{zhu2007real}, which helps reduce resonance effects during signal reconstruction.


\subsection{Pitch modification (PM)}
\label{subsec:PM}

Pitch modification was also implemented using the RTISI-LA algorithm, as described in \cite{2018explicit}, allowing adjustments in pitch based on a semitone value controlled by a scaling factor $\tau$, where $\tau$ varied between $0.5 \leq \tau \leq 2$ with a step size of 0.1. This method applies STFTM for processing the audio signal. The initial frame length and hop size are set, and the semitone value is converted into a corresponding scaling factor. The algorithm iterates through each frame of the input signal, applying a modified window function and reconstructing the signal iteratively. The window function is dynamically adapted according to the adjusted frame length, influenced by the semitone value, to achieve the desired pitch modification. This approach offers flexibility in modifying the tonal characteristics of the audio output. The semitone value has a direct impact on the resampling process during iterative reconstruction, playing a key role in accurate pitch adjustments \cite{zheng2017relation}. The modified audio is subsequently used for augmentation.


\subsection{Formant modification (FM)}
\label{subsec:FM}

Formant modification \cite{formant_mod, kathania2022data} is a key technique in speech processing that alters the spectral characteristics of speech signal by adjusting its formant frequencies. This process is implemented through Linear Predictive Coding (LPC) models, fitted to short-time segments of speech. LPC analysis is a robust method for modeling the spectral envelope of speech signals \cite{johnson2022lpc}, representing the signal as the output of an all-pole filter. The coefficients of this filter provide insight into the resonance frequency properties of the speech. For formant modification, these LPC coefficients are adjusted to achieve desired changes in the spectral envelope.

In applications like modifying children's speech, as demonstrated in \cite{formant}, formant modification is achieved by warping the poles of the LPC model, which effectively adjusts the formants. The altered LPC coefficients are then used to resynthesize the speech. The main steps in the formant modification process in this study are as follows: LPC analysis was performed on short-time segments of the speech signal. Coefficients from LPC were then modified through pole warping, controlled by a factor $\alpha$, where $-1 \leq \alpha \leq 1$ with a step size of 0.05, to shift the formants. The modified signal was synthesized using the warped LPC coefficients. 

\subsection{Vocal tract length perturbation (VTLP)}
\label{subsec:VTLP}

VTLP is employed to simulate variations in vocal tract length across different speakers, building on the concept of Vocal Tract Length Normalization (VTLN)~\cite{vtln_ref1}. While VTLN normalizes speaker variability, VTLP introduces controlled perturbations to the vocal tract length, thereby creating variability in the speech data. This method transforms a time-domain audio signal x(t) into the frequency domain as X(f) using the Fourier transform. A perturbation factor $\beta$ is then applied to adjust the frequency axis, where Y(f)=X($\beta f$). Values of $\beta < 1$ simulate a shorter vocal tract, while $\beta > 1$ simulates a longer one. This perturbation alters the spectral envelope while preserving the original duration of the audio.

In our experiments, the perturbation factor $\beta$ ranged between 0.98 and 1.08 with a step size of 0.02, introducing variability specifically targeting the spectral characteristics associated with different vocal tract lengths. This method effectively enhances the diversity of speech data, crucial for handling inter-speaker variability in speech recognition tasks.

\section{Experimental setup}
\label{sec:expsetup}

\subsection{Database description}
\label{ssec:feature}
The TORGO \cite{torgo} database, a widely used English dysarthric speech corpus, was utilized for the experiments in this study. The TORGO corpus contains approximately 15 hours of data from 15 speakers, including 8 dysarthric speakers and 7 age-matched healthy controls. All 15 speakers were included in our study. Data pre-processing followed the methodology outlined in \cite{decode_limit}, which involved removing empty audio files, files with complex transcripts, and files without transcriptions. The severity levels of the dysarthric speakers, evaluated using the Franchey Dysarthria Assessment (FDA), are based on speech intelligibility and are provided in Table \ref{tab:severity_labelled}.

\renewcommand{\arraystretch}{1.5}
\begin{table}[h]
    \centering
    \caption{Severity level description of dysarthric speakers in TORGO corpus.}
    \label{tab:severity_labelled}
    \resizebox{0.48\textwidth}{!}{%
    \begin{tabular}{|c|c|c|c|}\hline
        Corpus & Low & Medium & High  \\\hline
        TORGO & F03, F04, M03 & F01, M05 & M01, M02, M04  \\\hline

    \end{tabular}
}    
\end{table} 

\subsection{Proposed Framework }
\label{ssec:proposed_framework}
\begin{figure*}[h]
   \centering
   \includegraphics[width=\linewidth]{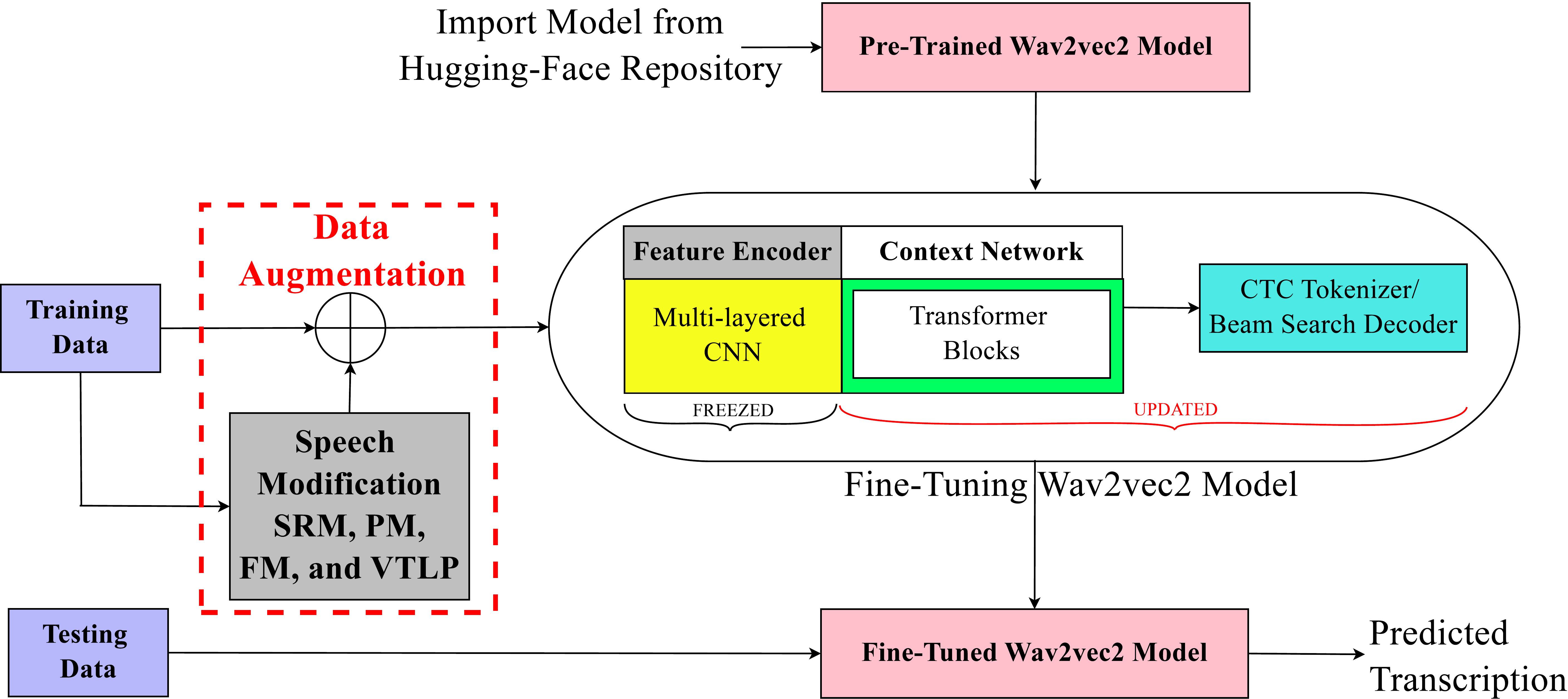}
   \caption{Block diagram of the proposed framework for employing data augmentation during the fine-tuning of a pre-trained Wav2Vec2 model.}
   \label{fig:asilomar}
\end{figure*}

The proposed approach for implementing data augmentation during the fine-tuning of the Wav2Vec2 model on the dysarthric speech recognition task is illustrated in Figure \ref{fig:asilomar}.

Initially, we imported the Wav2Vec2 model from the Hugging Face repository \cite{huggingface}. The large pre-trained variant, \textit{wav2vec2-large-960h-lv60-self}, was utilized in our study. This variant was trained on 60K hours of unlabeled data from the Libri-Light corpus \cite{librilight} and fine-tuned on 960 hours of labeled Librispeech \cite{panayotov2015librispeech} data, using a contrastive learning approach. A key advantage of this model is the incorporation of a self-training component \cite{liu2023towards}, which enhances adaptability for ASR tasks. Data augmentations were applied exclusively to the training set to ensure the test set remained unchanged, enabling easier comparison of results. The word error rate (WER) metric was used to evaluate ASR performance.

\subsection{Fine-tuning system configuration}
\label{ssec:ASR}

All experiments were conducted on a dedicated A5000 16GiB desktop GPU running Ubuntu. The experimental setup involved fine-tuning the Wav2Vec2 model, specifically the variant selected from baseline experiments, on the TORGO dysarthric speech dataset. The model was initialized with pre-trained weights using the Wav2Vec2ForCTC class from the Transformers library, with the feature encoder frozen to preserve the pre-trained representations. CTC loss with mean reduction \cite{connectionist} was applied during training. 

Training configurations included grouping the dataset by length for efficiency, with batch sizes set to 4 for training and 2 for evaluation. The training spanned 50 epochs, utilizing mixed-precision training (fp16) \cite{zhuang2023survey} and gradient checkpointing \cite{liu2020understanding} to manage memory. Checkpoints were saved every 1000 steps, and a learning rate of 0.0001 with a weight decay of 0.005 was employed, alongside a warm-up strategy over 2000 steps.  The training process, managed by the Trainer class from the Transformers library, incorporated the model, data collator, training arguments, and evaluation metrics, with WER as the primary evaluation metric. Severity-independent training and testing were conducted on the TORGO dataset.
\section{RESULTS and Discussion}
\label{sec:result}

Experiments were conducted by fine-tuning on speech from one severity level and testing on subsequent severity levels. To evaluate the baseline performance, the Wav2Vec2 model was fine-tuned using data from a single severity level in each experiment. The baseline results are provided in Table \ref{tab:baseline}.

\begin{table}[h]
\centering
\caption{Baseline fine-tuning results per severity level, in Word Error Rate (WER).}
\label{tab:baseline}

\begin{tabular}{|c|c|c|c|}\hline
    \multirow{2}{*}{Train-severity} & \multicolumn{3}{c|}{Test-severity} \\ \cline{2-4}
    &Low&Medium&High \\ \hline
    Low&--&80.13 &79.38 \\ 
    Medium&14.94&--&\bf{65.24} \\ 
    High&\bf{12.89}&\bf{45.72}&-- \\ \hline
\end{tabular}
\end{table}

When the model was trained on high-severity speech, the WER for low-severity speech was 12.89\%, and for medium-severity speech, it was 45.72\%. Similarly, when trained on medium-severity speech, the WER for high-severity speech was 65.24\%. The average of these scores resulted in a WER of 41.28\%, which serves as the baseline for comparison with results obtained through data augmentation.

\begin{figure}[h]
   \centering
   \includegraphics[width=1.2\linewidth]{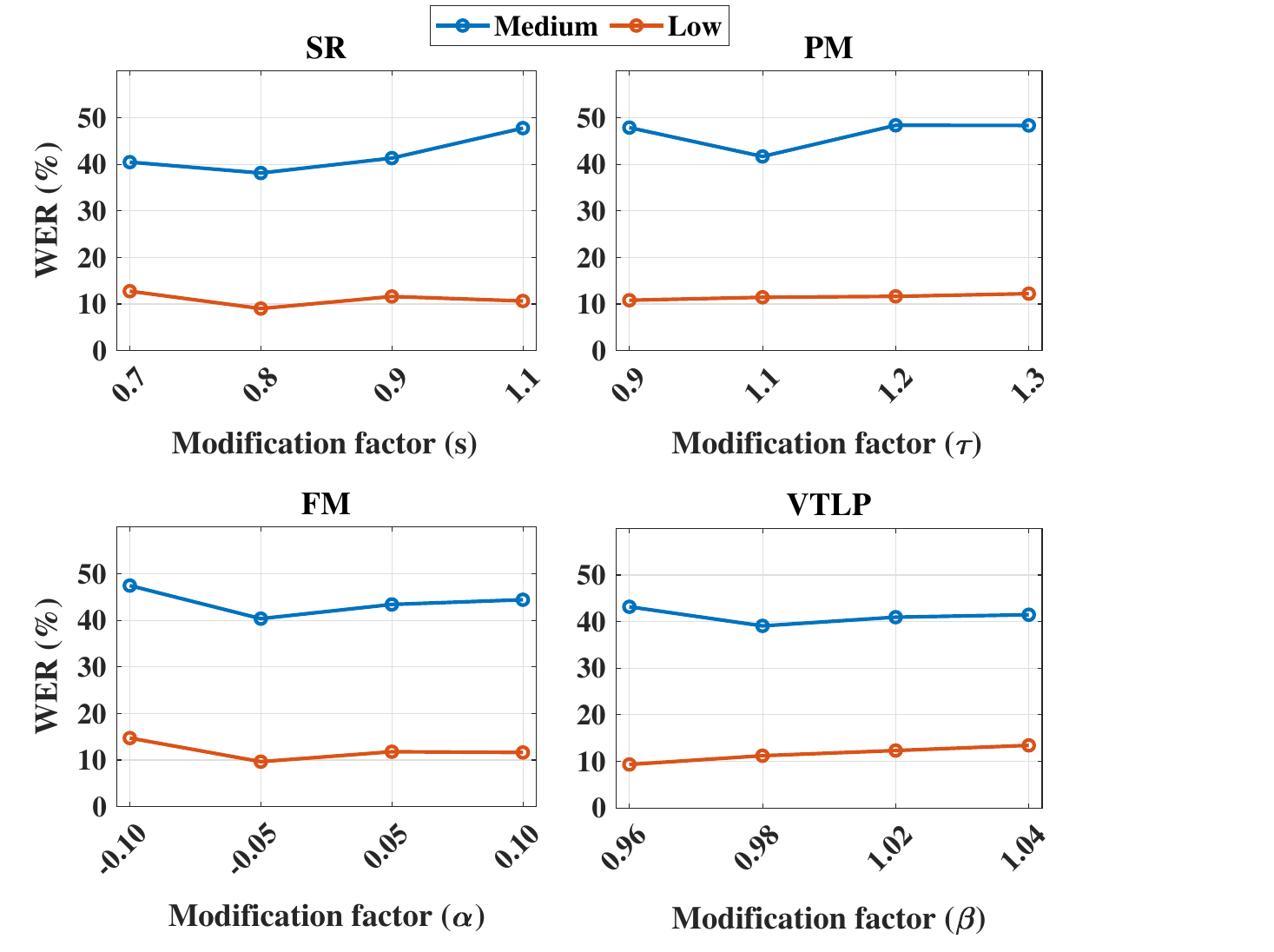}
   \caption{Results for the augmentation of \textit{high} severity speakers' data during training/fine-tuning and testing with \textit{low} and \textit{medium} severity levels, with Word Error Rate (WER) as the evaluation metric. The values of $s$, $\tau$, $\alpha$, and $\beta$ denote the modification factors for SRM, PM, FM, and VTLP-based speech data augmentations.}
   \label{fig:augmentation_high}
\end{figure}

Data augmentation was applied by varying the modification factors ($s$, $\tau$, $\alpha$, and $\beta$) to enhance the training data. For each severity level, the training set was augmented accordingly. Figure \ref{fig:augmentation_high} illustrates the results of data augmentation when training/fine-tuning on speakers with \textit{high} severity and testing with \textit{low} and \textit{medium} severity levels, highlighting the performance across different variations of the modification factors.

Figure \ref{fig:augmentation_high} shows that reducing the SRM modification factor $s$ significantly improves ASR performance for both \textit{medium} and \textit{low} severities. This trend sustained till $s=0.8$, following which the WERs started increasing. For PM, the best result for \textit{medium} severity was observed at $\tau=1.1$, while a consistent improvement trend was noted for \textit{low} severity as $\tau$ decreased. Similarly, FM achieved the lowest WER at $\alpha=-0.05$. In the case of VTLP, reducing the modification factor $\beta$ improved performance for \textit{low} severity, with a significant enhancement at $\beta=0.98$ for \textit{medium} severity speech.

\begin{figure}
   \centering
   \hspace{-0.5cm}
   \includegraphics[width=1.2\linewidth]{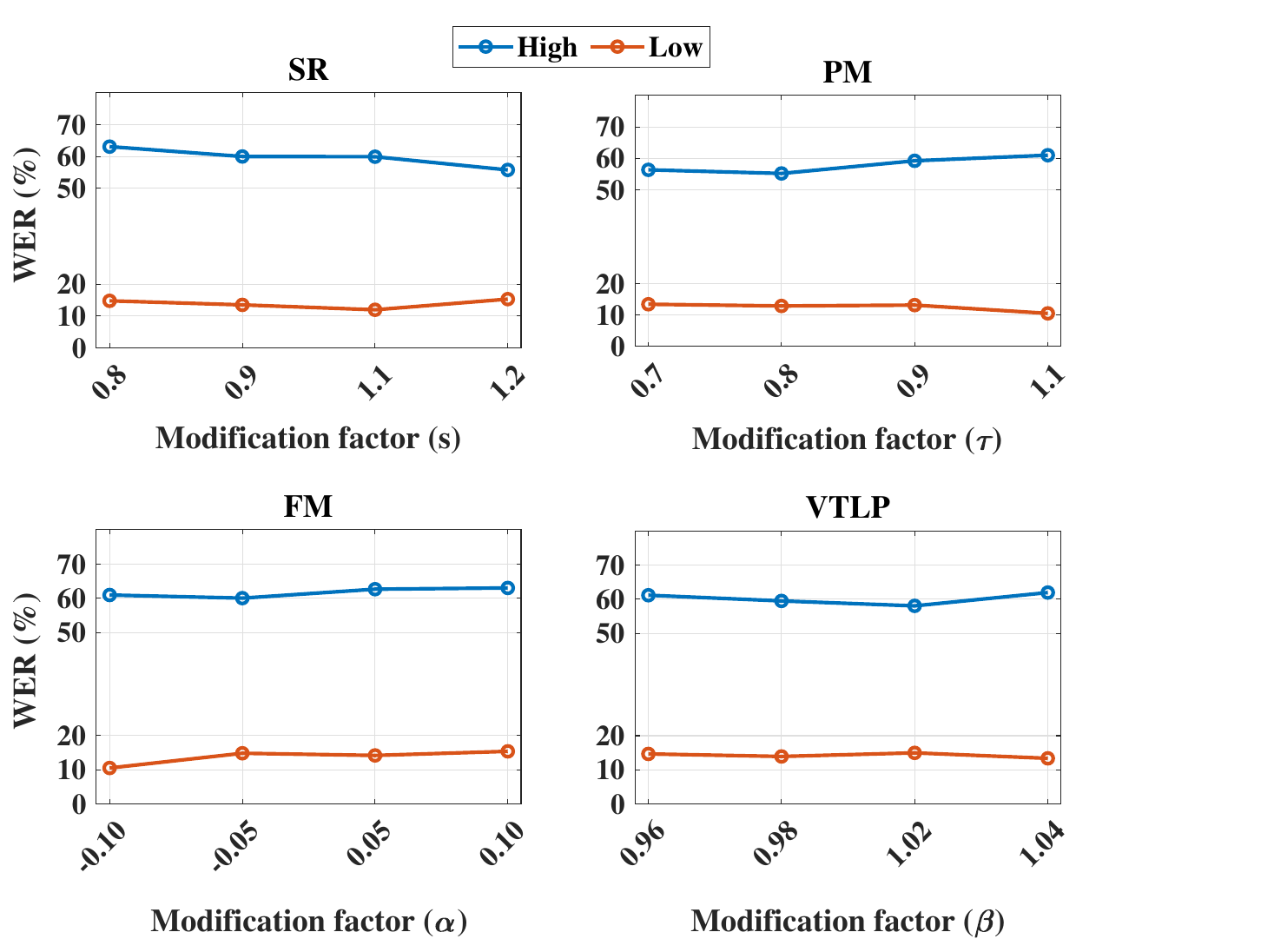}
   \caption{Results for the augmentation of \textit{medium} severity speakers' data during training/fine-tuning and testing with \textit{low} and \textit{high} severity levels, with Word Error Rate (WER) as evaluation metric. The values of $s, \tau, \alpha$, and $\beta$ denote the modification factors for SRM, PM, FM, and VTLP based speech data-augmentation.}
   \label{fig:augmentation_medium}
\end{figure}

The fine-tuning results of the \textit{medium} severity model on the test sets of \textit{low} and \textit{high} severities are illustrated in Figure \ref{fig:augmentation_medium}. Notably, substantial improvements were achieved when data augmentation techniques were applied to the evaluation of \textit{high} severity dysarthric speech. As seen in the figure, for the \textit{high} severity test case, SRM yielded the best results with a modification factor of $s=1.2$ for \textit{high} severity, while PM performed optimally with $\tau=0.8$. Similarly, FM showed enhanced performance with $\alpha=-0.05$, and the most effective factor for VTLP was $\beta=1.02$. These observations are based on the WER from the \textit{high} severity evaluation. 

In Figure \ref{fig:augmentation_medium}, the evaluation of \textit{low} severity test data highlights the distinct performance of the four augmentation methods employed. The SRM and PM methods demonstrated optimal performance with a modification factor of $s$ and $\tau=1.1$.
FM exhibited a improved performance with $\alpha=-0.1$, and VTLP showed the better results with $\beta=1.04$. Among these, the FM augmentation method performed better than the others, delivering the most effective results for this evaluation.

Table \ref{tab:best_factor} presents the best modification factors for each of the four augmentation methods, providing a clear comparison of their effects on model performance. Fine-tuning/training with augmented \textit{high}-severity speech (for all four augmentation methods)  demonstrated the improved ASR performance when evaluating \textit{low} and \textit{medium}-severity speech. SRM with $s=0.8$ achieved the best WERs of 9.02\% and 38.11\% for \textit{low} and \textit{medium} severities, respectively. Fine-tuning with \textit{medium}-severity speech resulted in notable WER reductions when testing on \textit{high}-severity speech, with PM yielding the best performance at 55.15\%. Both relative and absolute improvements for each severity level are summarized in Table \ref{tab:best_factor}. Our approach of using data augmentation during fine-tuning of the Wav2Vec2 model achieved relative WER improvements of 30.02\%, 16.64\%, and 15.47\% for \textit{low}, \textit{medium}, and \textit{high} severities, respectively, compared to the baseline fine-tuning results.

\renewcommand{\arraystretch}{1.3}
\begin{table}[h]
    \centering
    \caption{Word Error Rates (WERs) after applying data augmentation with the optimal modification factors: $s$, $\tau$, $\alpha$, and $\beta$. The subscripts $_L$', $_M$', and `$_H$' represent the factors used for \textit{low}, \textit{medium}, and \textit{high} severity levels, respectively.}
    \label{tab:best_factor}
    \begin{tabular}{|c|c|c|c|c|}\hline
         Train-Severity&Augmentation Method & \multicolumn{3}{c|}{Test-Severity} \\\cline{3-5}
         &&Low&Medium&High \\ \hline\hline
         \multicolumn{2}{|c|}{Baseline best cases}&12.89&45.72&65.24 \\ \hline \hline
         
         \multirow{4}{*}{High}&SRM ($s$=0.8$_{L,M}$)&\bf{9.02}&\bf{38.11}&-- \\ \cline{2-5}
         &PM ($\tau$=0.9$_L$, 1.1$_M$)&10.81&41.66&-- \\ \cline{2-5}
         &FM ($\alpha$=-0.05$_{L,M}$)&9.65&40.34&-- \\ \cline{2-5}
         &VTLP ($\beta$=0.96$_L$, 0.98$_M$)&9.34&39.08&-- \\ 
         \hline
         \multirow{4}{*}{Medium}&SRM ($s$=1.1$_L$, 1.2$_H$)&11.95&--&55.80 \\ \cline{2-5}
         &PM ($\tau$=1.1$_L$, 0.8$_H$)&10.57&--&\bf{55.15} \\ \cline{2-5}
         &FM ($\alpha$=-0.1$_L$, -0.05$_H$)&10.48&--&60.05 \\ \cline{2-5}
         &VTLP ($\beta$=1.04$_L$, 1.02$_H$)&13.37&--&58.08 \\ \hline
         \hline
         \multicolumn{2}{|c|}{Best cases after augmentation}&9.02&38.11&55.15 \\ \hline \hline
         \multicolumn{2}{|c|}{Relative improvements in (\%)}&30.02&16.64&15.47 \\ \hline
         \multicolumn{2}{|c|}{Absolute improvements in (\%)}&3.87&7.61&10.09 \\ \hline
         
    \end{tabular}
    
\end{table}

 
Note that this study exclusively focused on the Wav2Vec2-based end-to-end ASR system. However, our other study has systematically explored the impact of acoustic models and features in traditional Kaldi-based ASR system, and readers can refer to \cite{Kladi_Dys_Asilomar}.

\section{CONCLUSIONS}
\label{sec:typestyle}

In this study, we evaluated the impact of various data augmentation methods on fine-tuning a Wav2Vec2-based ASR system for dysarthric speech. Among the four methods, speaking rate modification (SRM) proved most effective for decoding \textit{low} and \textit{medium} severity speech, while pitch modification (PM) was the most beneficial for recognizing \textit{high} severity speech. Compared to the baseline results without augmentation, each method significantly enhanced ASR performance. The lowest WERs were achieved with SRM ($s=0.8$), yielding 9.02\% for \textit{low} severity and 38.11\% for \textit{medium} severity, representing relative improvements of 30.02\% and 16.64\%, respectively. For \textit{high} severity, PM ($\tau=0.8$) achieved the best WER of 55.15\%, reflecting a 15.47\% relative improvement. Our findings underscore the importance of selecting the optimal modification factor based on the severity level of the test data. This study demonstrated that data augmentation methods effectively address data scarcity and enhances the ASR performance for dysarthric speech. Our findings also show that incorporating high-severity data during fine-tuning improves system performance across both low and medium severity levels, suggesting that this approach has the potential to generalize and improve dysarthric ASR.

A potential future direction involves addressing the challenge of unknown speaker severity in real-world applications. While this study used known severity labels for severity-specific experiments, practical deployment requires automatic severity classification before implementing severity-specific ASR systems. Future research could explore severity detection using SSL features, as in \cite{Farhad_BHI,farhad2023pretrained,farhad_specom}, and integrate this approach into the End-to-End ASR system to enhance performance.
\bibliographystyle{IEEEbib}
\bibliography{spcom,refs}

\end{document}